\documentclass[fleqn,twoside]{article}
\usepackage{espcrc2}
\usepackage{graphicx}
\usepackage[figuresright]{rotating}

\newcommand{\J}{J/\psi}
\newcommand{\psip}{\psi(2S)}
\newcommand{\EE}{e^+e^-}
\newcommand{\MM}{\mu^+\mu^-}
\newcommand{\TT}{\tau^+\tau^-}
\newcommand{\GG}{\gamma\gamma}
\newcommand{\pp}{\pi^+\pi^-}
\newcommand{\PPJP}{\pi^+\pi^-J/\psi}

\newcommand{\ra}{\rightarrow}

\title{ A Measurement of $\psi(2S)$ Resonance Parameters}
\author{\small 
\mbox{} \hskip 5cm (BES Collaboration) \\
\vspace{0.2cm}
J.~Z.~Bai$^1$, Y.~Ban$^{10}$,      J.~G.~Bian$^1$, I.Blum$^{18}$,
X.~Cai$^{1}$,       J.~F.~Chang$^1$,
H.~F.~Chen$^{17}$,  H.~S.~Chen$^1$,  J.~Chen$^{4}$,
Jie~Chen$^{9}$
% ...
\\
\vspace{0.1cm}
$^1$ Institute of High Energy Physics, Beijing 100039, People's Republic of
     China\\
% ...
$^4$ Colorado State University, Fort Collins, Colorado 80523\\
% ...
$^9$ Nankai University, Tianjin 300071, People's Republic of China\\
$^{10}$ Peking University, Beijing 100871, People's Republic of China\\
% ...
$^{17}$ University of Science and Technology of China, Hefei 230026,
        People's Republic of China\\
$^{18}$ University of Texas at Dallas, Richardson, Texas 75083-0688\\
% ...
}

\begin{document}
\begin{abstract}
Cross sections for $\EE \rightarrow$ hadrons, $\PPJP$, and $\MM$ have been
measured in the vicinity of the $\psip$ resonance using the BESII detector
operated at the BEPC. The $\psip$ total width; partial widths to
hadrons, $\PPJP$, leptons; and corresponding branching fractions 
have been determined to be
$\Gamma_t=264 \pm 27 $ keV;
$\Gamma_h=258 \pm 26 $ keV,
$\Gamma_{\PPJP}=85.4 \pm 8.7 $ keV, and
$\Gamma_{l}=2.44 \pm 0.21$ keV; and
$B_h=(97.79 \pm 0.15) \%$,
$B_{\PPJP}=(32.3\pm 1.4)\%$, and
$B_{l}=(0.93\pm 0.08) \%$, respectively.
\vspace{1pc}
\end{abstract}
\maketitle

\normalsize
\section{Introduction}
\indent 
   Since the discovery of the $\psip$ in 1974~\cite{ref.1}, 
measurements of its total width ($\Gamma_t$), and partial decay widths
into hadrons ($\Gamma_h$), $\PPJP$ ($\Gamma_{\PPJP}$), and $l^+l^-$ 
($\Gamma_l$), and the corresponding branching fractions, 
$B_h$, $B_{\PPJP}$, and $B_l$, have been carried out [2$-$10].
The results of these experiments differ on both decay widths and
branching fractions. 
The parameters are of particular interest because, for instance, 
$\psip\rightarrow l^+ l^-$ is used in reconstructing $B$ mesons 
for $CP$ violation measurements \cite{babar}, 
and $\psip\rightarrow\PPJP$ is often used to 
determine the total number of $\psip$ events in $\psip$ branching
fraction measurements due to its large branching fraction and 
straightforward detection. Therefore, it is important to measure 
these decay widths and branching fractions with better accuracy.

Twenty four center-of-mass energy points were 
scanned in the vicinity of
the $\psip$ peak ranging from 3.67 GeV to 3.71 GeV.
The data were collected with the BESII (BEijing Spectrometer)
detector at the BEPC (Beijing Electron Positron Collider)
storage ring. The BESII detector is described in detail in 
Ref.~\cite{bes.1}. In addition, separated-beam data were taken at the first 
and the last points for background studies.
 The total 
integrated luminosity was 1149 nb$\mbox{}^{-1}$. 

\section{Event selection}\label{sect_two}
The following four reactions are studied:
\begin{eqnarray*}
~~~~~~~~~~~~~~~~~~\EE &\rightarrow& \EE ~~~,\label{reac.1} \\
\EE &\rightarrow& \MM ~~~,\label{reac.2} \\
\EE &\rightarrow& \pi^+ \pi^- J/\psi ~~~,\label{reac.3} \\
\EE &\rightarrow& \mbox{ hadrons} ~~~.\label{reac.4}
\end{eqnarray*}

\begin{figure}[htb]
\center
\includegraphics[height=5.50 cm,width=5.50cm ]{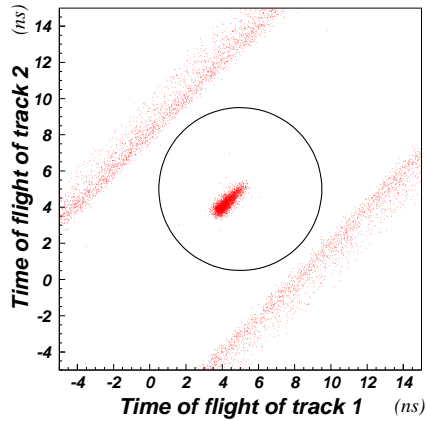}
\caption{\label{mutof} Times of flight for $\mu$-pairs. The
diagonal bands  are due to cosmic ray events.  Good events, which
cluster at the center of the plot are selected with the circle cut shown.}
\end{figure}

\noindent For the selection of lepton-pair final states, two charged tracks
with total charge zero are required. For $\MM$
events,
the acollinearity must be less than 10 degrees.
In addition, in order to suppress cosmic ray background, the time-of-flight
measurements of the two muon-candidates must satisfy
$\sqrt{(t_1-5)^2 + (t_2-5)^2} <4.5 $(ns), as shown in Fig.~\ref{mutof}.
Further, the Muon Counter hit information is used
to identify di-muon events from other back-to-back two-prong final states,
and this requires $|\cos \theta_{\mu}| \leq 0.65$ due to the limited
solid angle coverage.

To separate electrons from muons and hadrons, 
the energies deposited by the two tracks in the Barrel Shower Counter (BSC) 
must satisfy 
$\sqrt{(\tilde{E}_{dep1}-1)^2 + (\tilde{E}_{dep2}-1)^2} <0.65 $, where
$ \tilde{E}_{dep}=\frac{E_{dep}}{E_{beam}}$ is the normalized energy deposited.
We also require
$|\cos \theta_{e}| \leq 0.72 $, and
because the Monte Carlo simulation does not model the energy 
deposited well in the rib region of the BSC, an additional cut is 
applied on the $z$-coordinate of the first hit layer:
$ 0.03<|z_{sc}|<0.85$ or $|z_{sc}|>0.95$ m.

\begin{figure}[hbt]
\center
\mbox{}
\vskip 0.5cm 
\includegraphics[height=5.50cm,width=5.50cm]{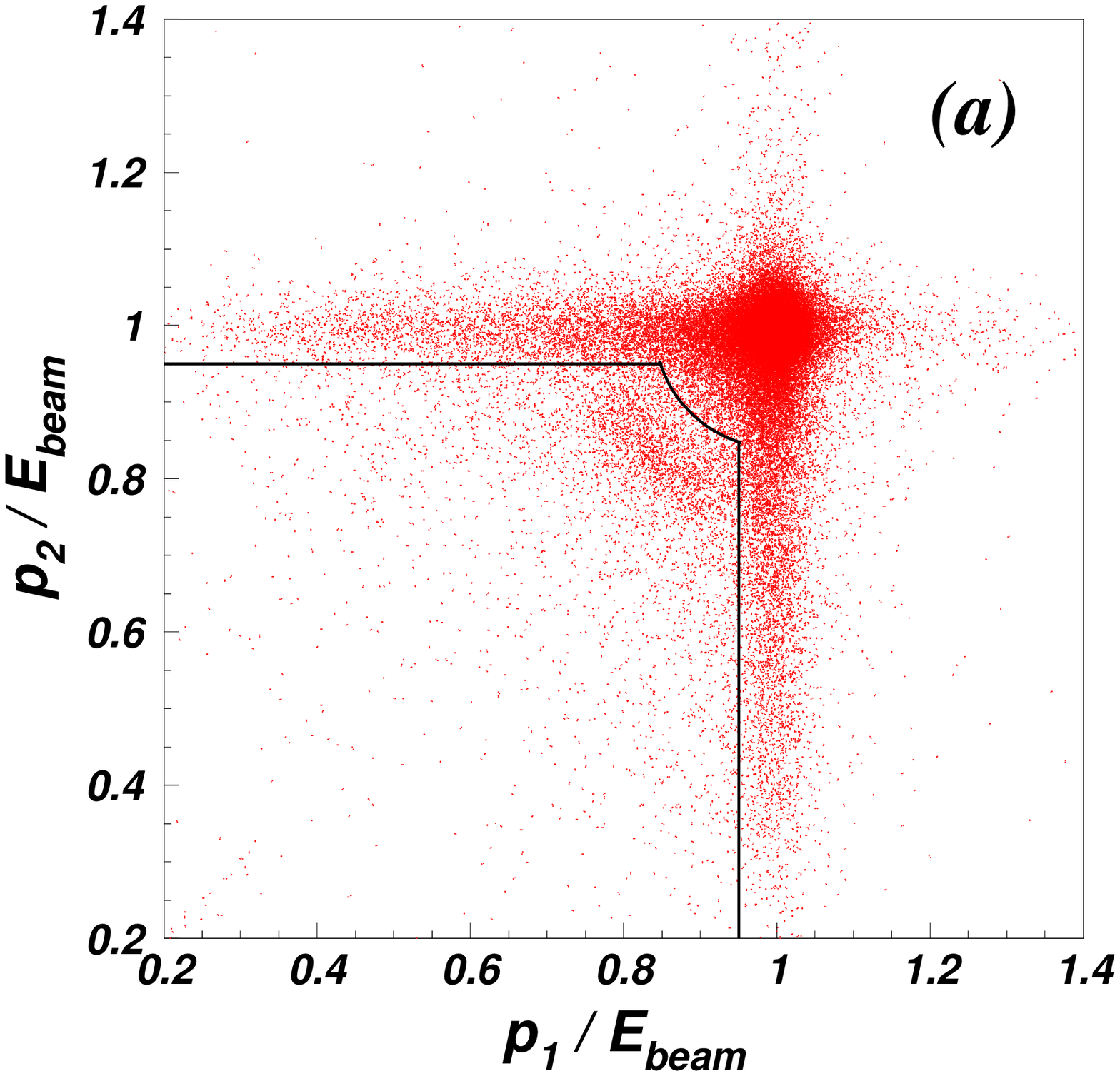}
\vskip 0.5 cm
\includegraphics[height=5.50cm,width=5.50cm]{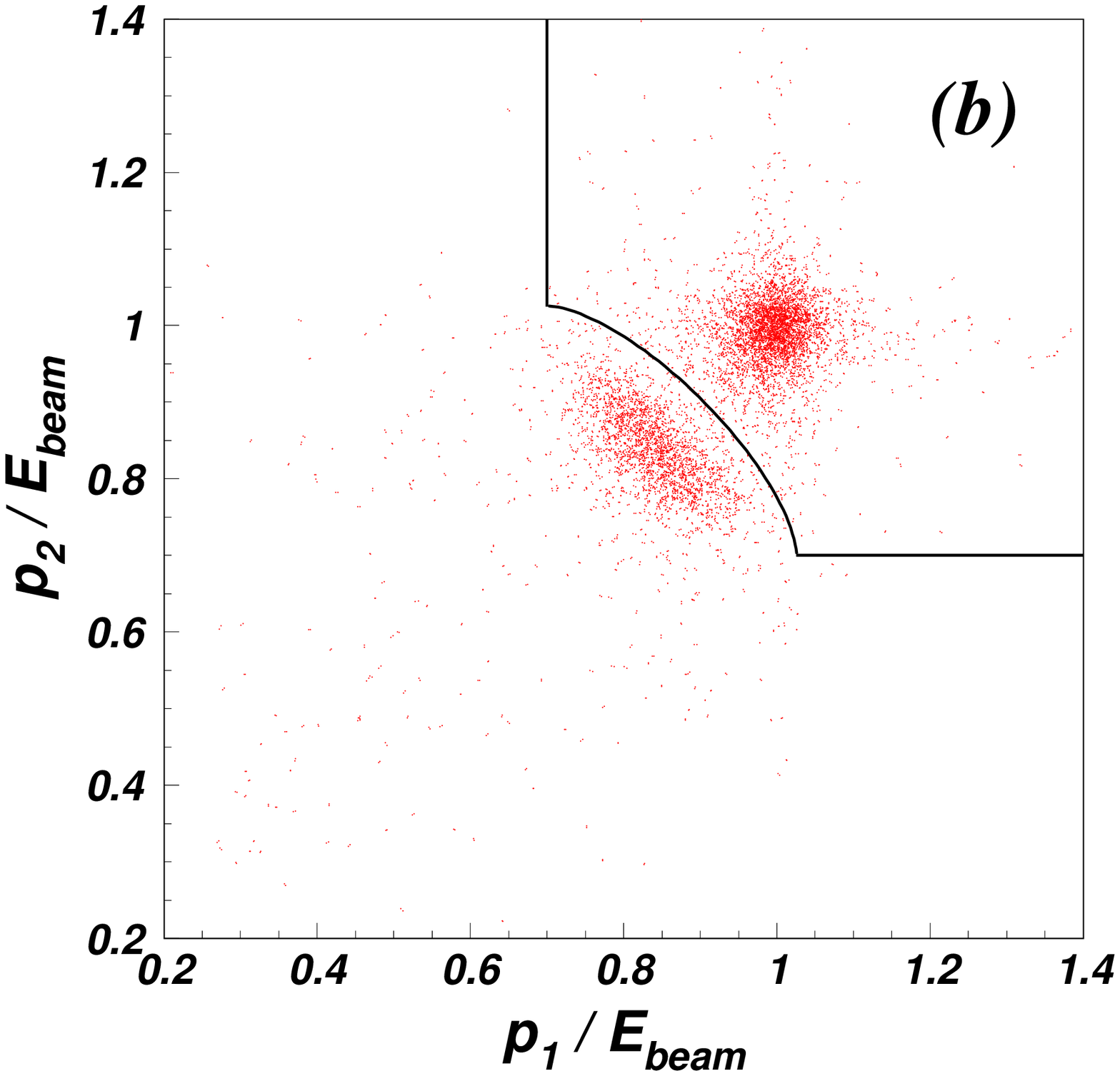}
\caption{\label{mixpcut} Momentum distributions for (a) $\EE$ events and
(b) $\MM$ events.  
The solid line indicates the cuts applied to remove background from
$\psip \rightarrow X J/\psi$, $J/\psi \rightarrow l^{+}l^{-}$ \cite{track_cuts}. }
\end{figure}

\begin{figure}[htb]
\center
\includegraphics[height=3.50cm ,width=5.50cm]{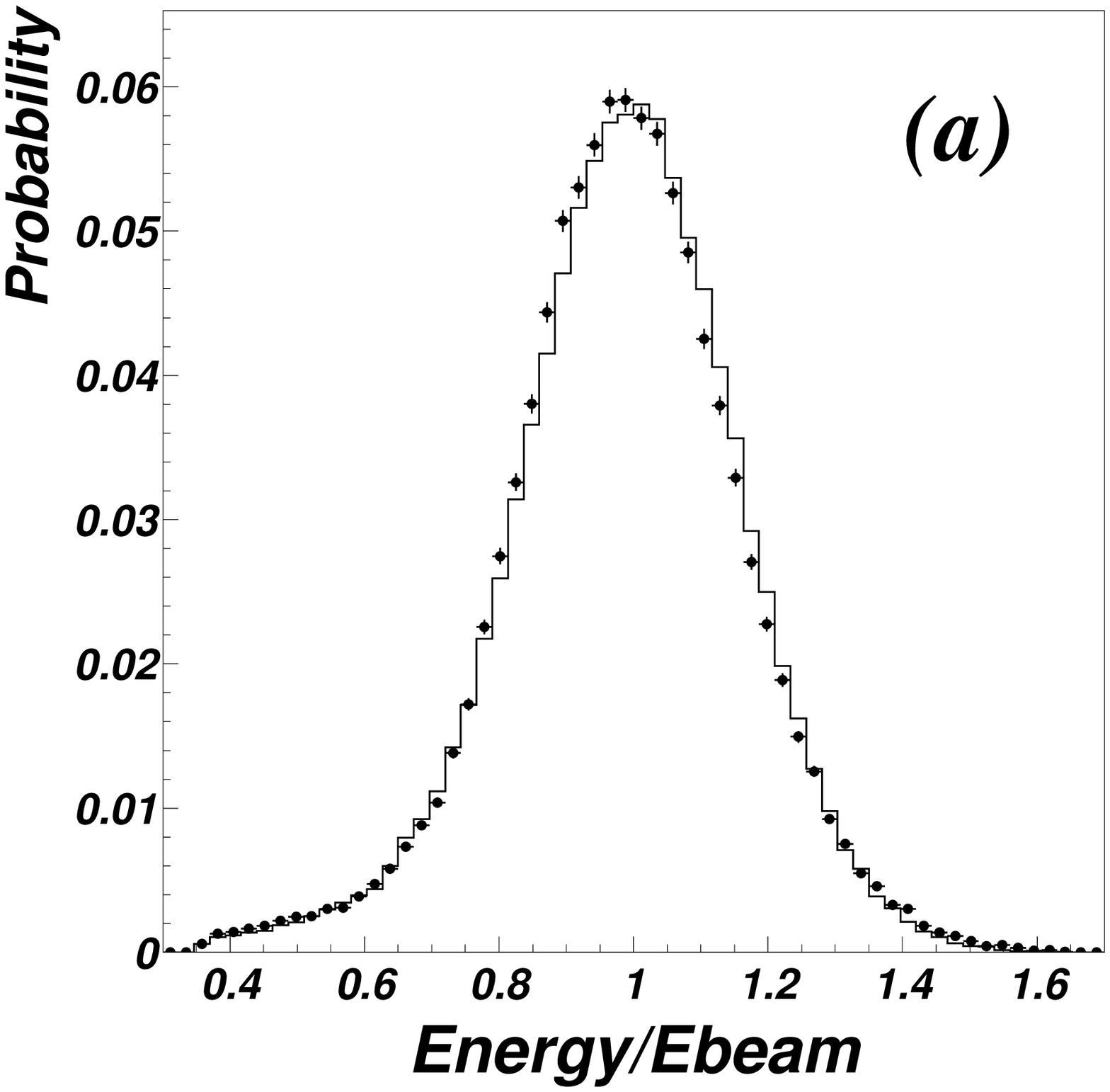}
\vskip 0.35 cm
\includegraphics[height=3.50cm ,width=5.50cm]{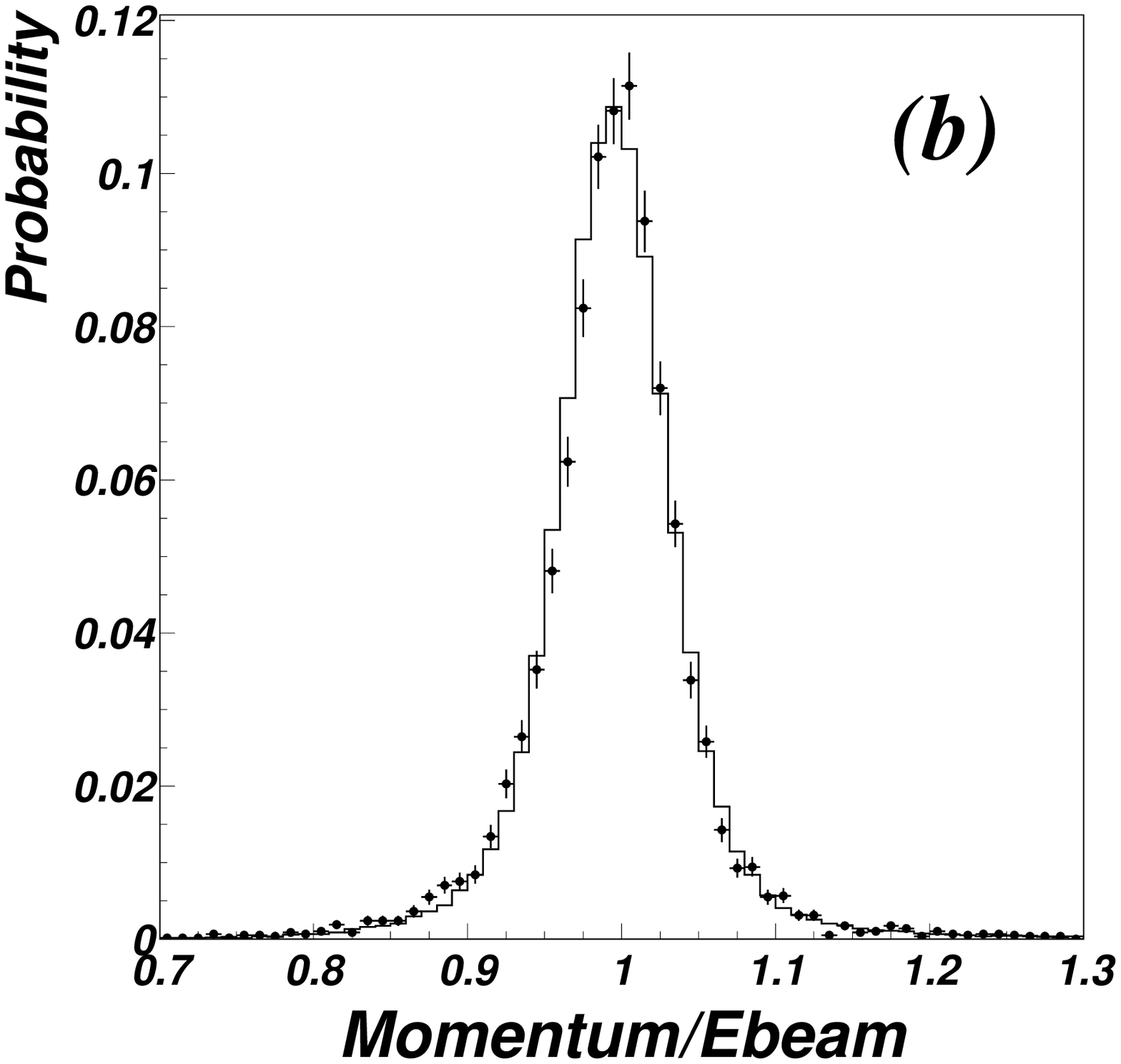}
\caption{\label{mixcmp}
Normalized momentum distributions for (a) $\EE$ events and 
(b) $\MM$ events.
(Histogram for Monte Carlo events and dots with error bar for data)}
\end{figure}

\begin{figure}[htb]
\center
\includegraphics[height=2.50cm ,width=6.0cm]{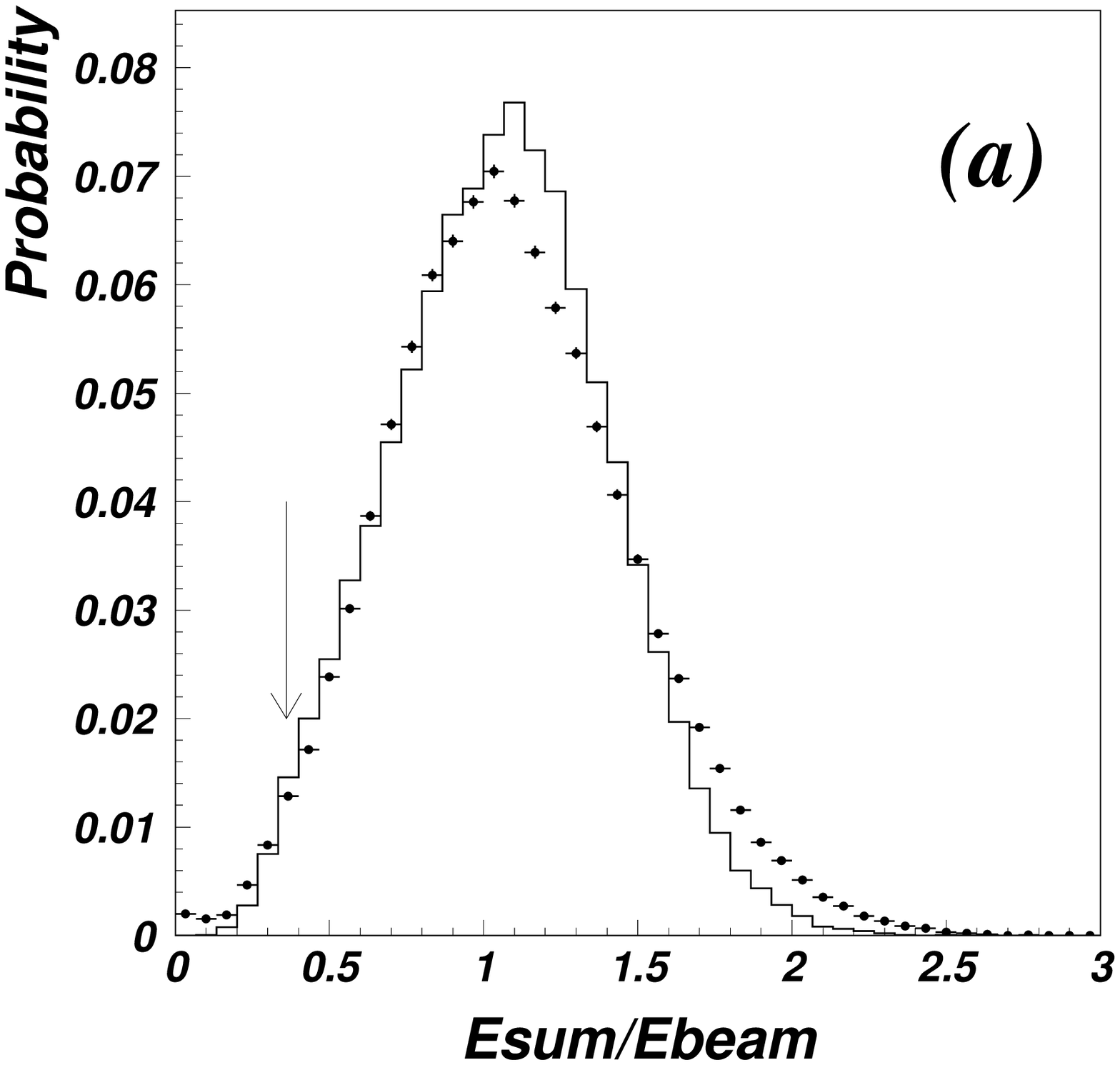}
\vskip 0.25 cm
\includegraphics[height=2.50cm ,width=6.0cm]{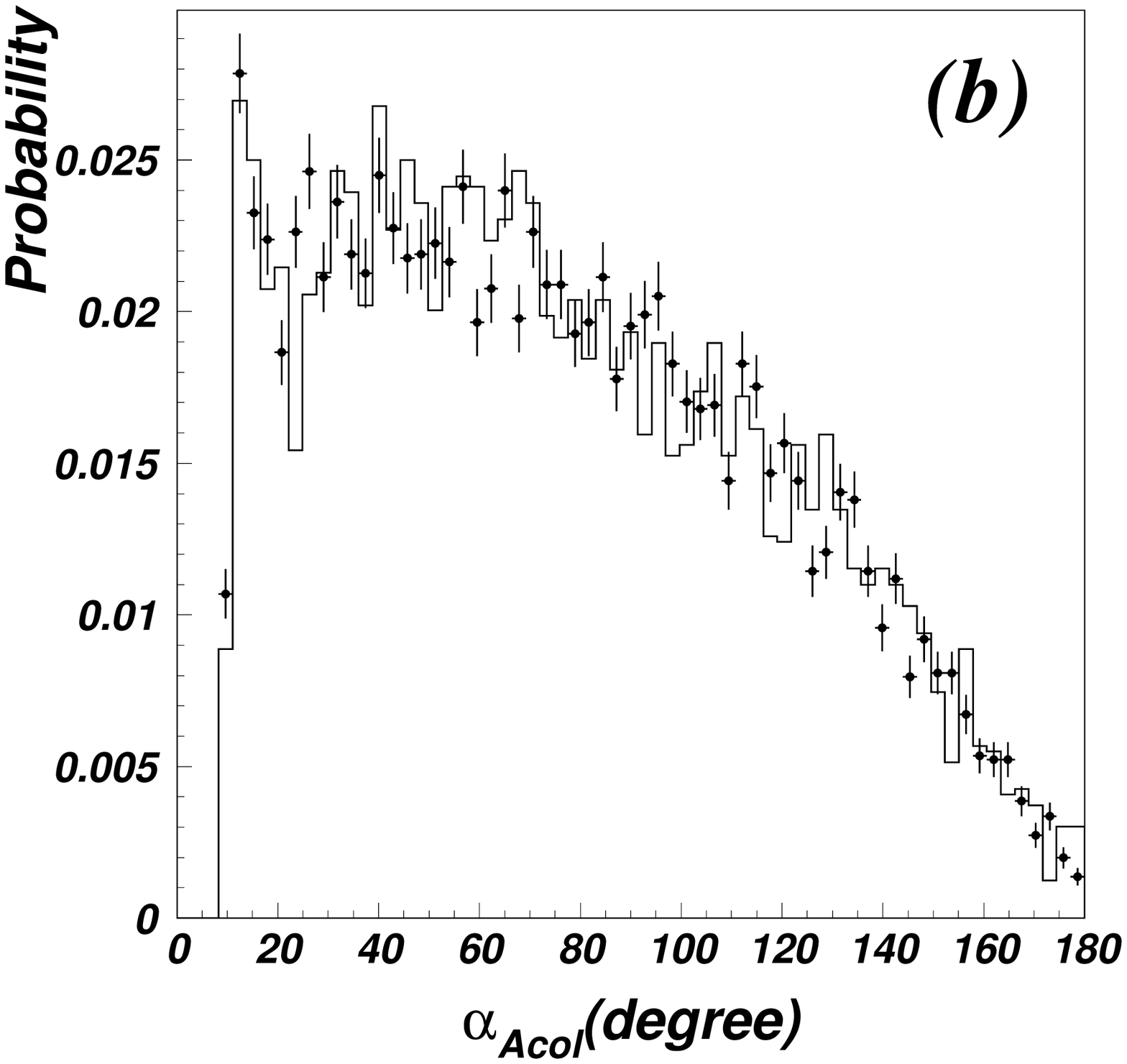}
\vskip 0.25 cm
\includegraphics[height=2.50cm ,width=6.0cm]{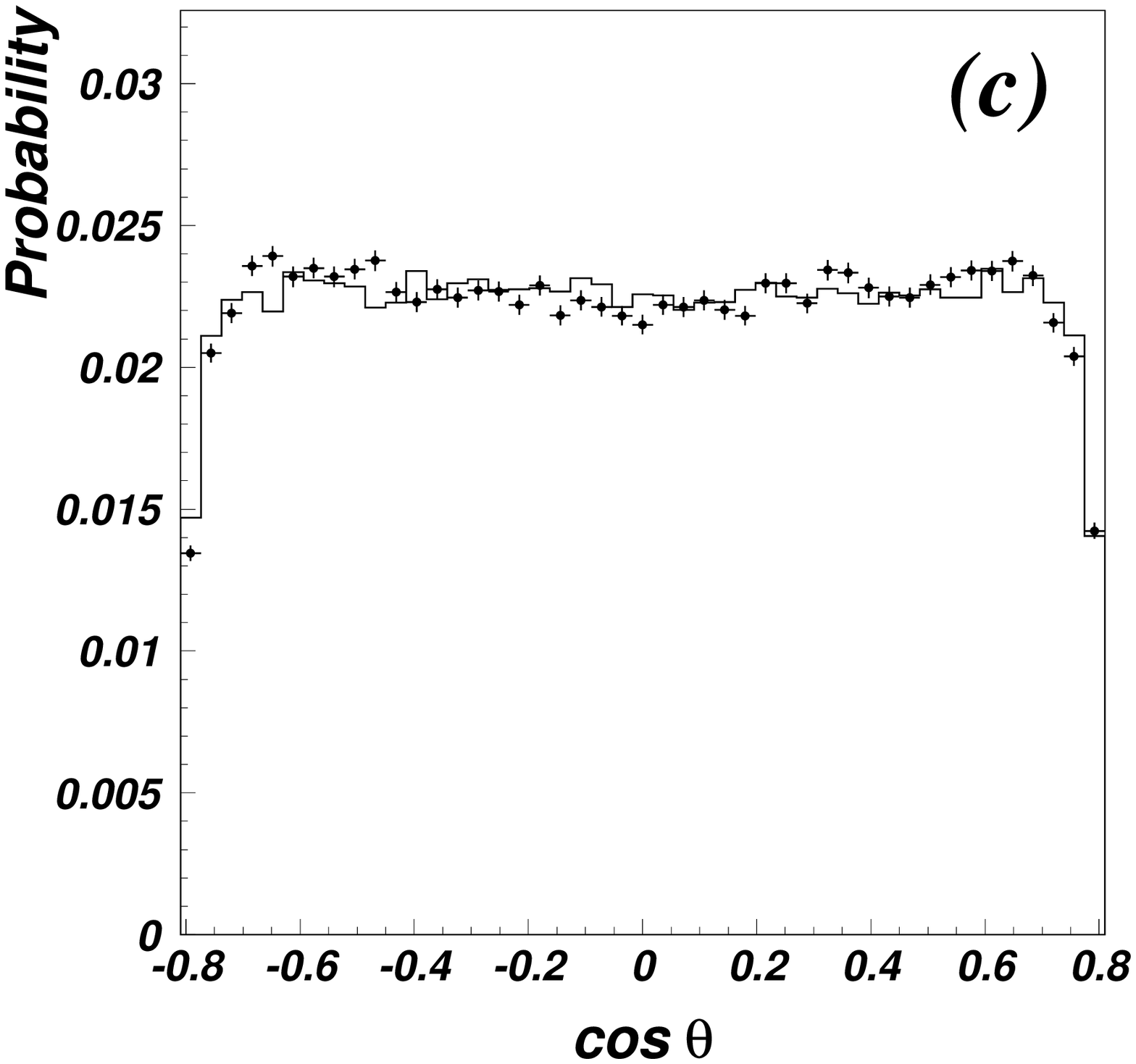}
\caption{\label{hadcmp}
Comparison between Monte Carlo events (histogram) and 
data (dots with error bar) for hadron events. 
(a) $Esum$ (b) $\alpha_{Acol}$ (c) $\cos \theta$ of charged track.
(The arrow in (a) indicates the cut applied.) }
\end{figure}

A potential source of background is lepton-pairs coming from
$\psip \rightarrow X J/\psi$, $J/\psi \rightarrow l^{+}l^{-}$.
To eliminate this background, we make cuts on the
track momenta, as shown in
Fig.~\ref{mixpcut} \cite{track_cuts}. 

\begin{figure}[hbt]
\center
\mbox{}
\vskip 0.3cm
\includegraphics[height=6.0cm,width=6.5cm]{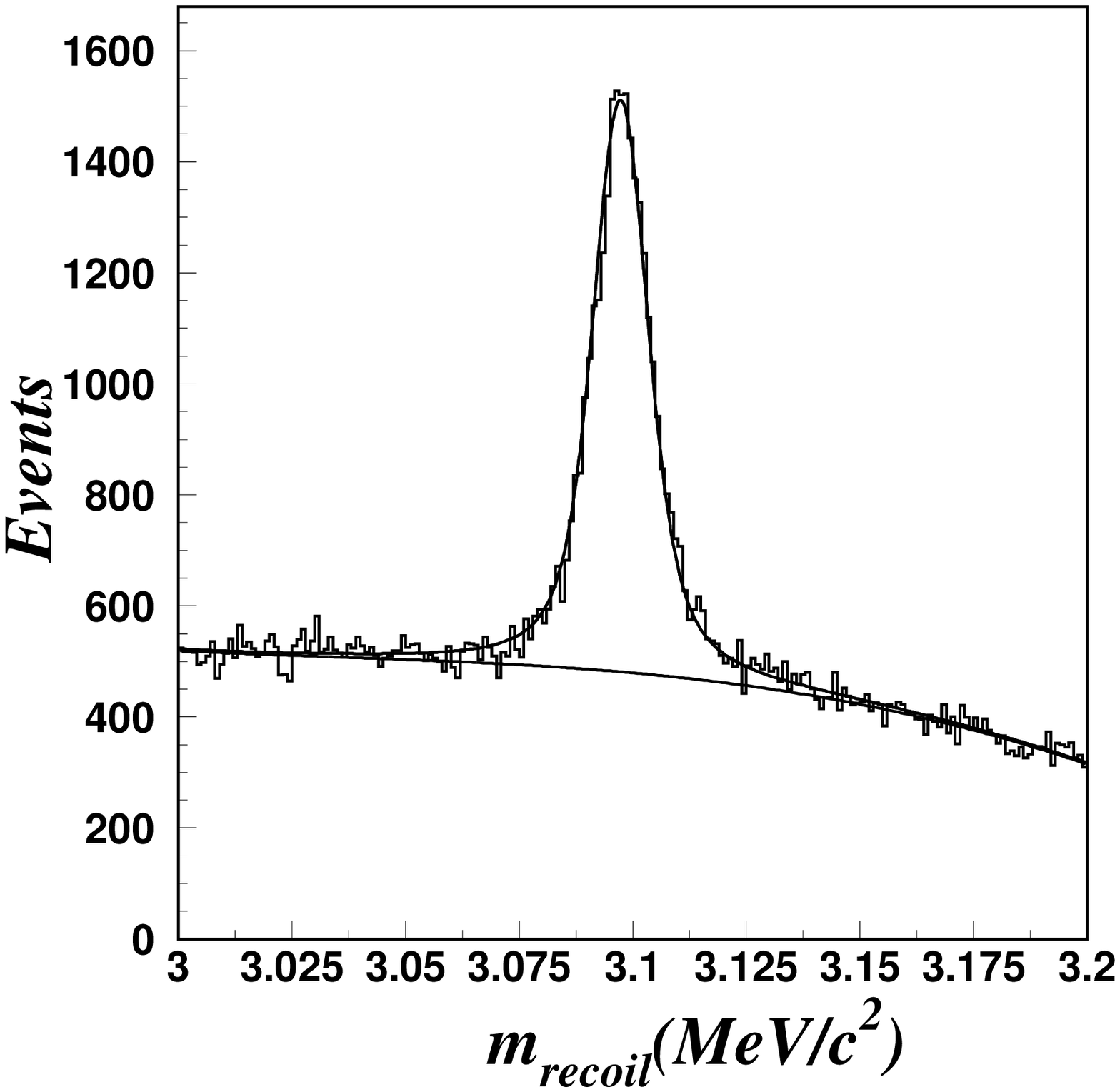}
\caption{\label{m_recoil_incl}
Mass recoiling against $\pp$ for all 24 energy points combined. The
$\J$ peak is due to the signal $\psip\ra\PPJP, \J\ra$anything events.} 
\end{figure}

Using the above criteria to select 
$\MM$ and $\EE$ events, we have compared  
various distributions with Monte Carlo 
distributions. Good agreement is found, as illustrated, for example, in 
Fig.~\ref{mixcmp}.

For hadron event selection, we are guided by our R-scan experience
\cite{trig,R-scan}. There is no particular event topology to require;
instead we make cuts to reject major backgrounds: cosmic rays,
beam-associated background, two-photon processes
($\gamma^{\ast}\gamma^{\ast}$), mis-identified ``hadron'' events from
QED processes of $\EE \rightarrow l^+l^-$, $l=e$, $\mu$, $\tau$, and $\EE
\rightarrow \GG$ followed by $\gamma$ conversion, etc.  Events
with at least two well reconstructed charged tracks within 
$|\cos \theta| \leq 0.8$ are selected. The total energy deposited by 
an event in the BSC ($E_{sum}$) is required to be larger
than $0.36 E_{beam}$, in order to suppress contamination from
two-photon processes and beam associated background. Events with all 
tracks pointing to the same hemisphere in the $z$ direction are removed 
to suppress beam-associated background.  For two-prong events, 
two additional cuts are applied to eliminate possible lepton pair background.
The number of photons must be greater than one, and the acollinearity 
between two charged tracks, $\alpha_{Acol}$,  
must be greater than 10 degrees. A comparison between 
data and Monte Carlo samples satisfied the foregoing selection criteria
is shown in Fig. \ref{hadcmp}.

The background from
$\TT$ decay is difficult to distinguish from direct hadronic decay
events, so the contribution from this source, $N_{\TT}$, is estimated
using $N_{\tau \tau}=L \cdot (\varepsilon_{\tau \tau} \cdot
\sigma_{\tau \tau}),$ where $L$ is the integrated luminosity at each
energy point, $\sigma_{\tau\tau}$ the QED production cross section at
this energy point, and $\varepsilon_{\tau \tau}$ the acceptance of our
hadron event selection criteria for $\TT$ events. This is subtracted
from the observed number of hadron events. A similar subtraction is
performed for the other surviving backgrounds, such as $\EE
\rightarrow e^+e^-$, $\mu^+\mu^-$, $\gamma \gamma$, and the two-photon
process ($\gamma^{\ast}\gamma^{\ast}$).  Therefore, the corrected
number of hadron events, $N_h^{obs}$, is
$$
N_h^{obs}=N_h-N_{\tau\tau}-N_{ee}-N_{\mu\mu}-N_{\gamma\gamma}-
   N_{\gamma^*\gamma^*} $$
where $N_h$ is the number of events that satisfy the hadron event selection 
cuts. The numbers of various backgrounds are given 
in Table \ref{hadbg}.

\begin{table}[htb]
\doublerulesep 0.5pt
\caption{\label{hadbg} Estimated numbers of hadron event backgrounds. }
\vskip 0.2 cm
\begin{center}
\begin{tabular}{c|ccccc}  \hline \hline
$E_{cm}$&\multicolumn{5}{|c}{ Background events } \\ \cline{2-6}
(GeV)&$N_{\tau\tau}$
           &$N_{ee}$&$N_{\mu\mu}$
	                 &$N_{\gamma\gamma}$
                                &$N_{\gamma^*\gamma^*}$ \\ \hline \hline
3.6668 & 20.79& 12.16&  0.22&  0.80&  1.45 \\
3.6719 & 20.36& 11.91&  0.21&  0.78&  1.42 \\
3.6750 & 20.15& 11.79&  0.21&  0.78&  1.41 \\
3.6781 & 20.72& 12.12&  0.22&  0.80&  1.46 \\
3.6801 & 20.58& 12.04&  0.21&  0.79&  1.45 \\
3.6809 & 22.18& 12.98&  0.23&  0.85&  1.56 \\
3.6820 & 20.80& 12.17&  0.22&  0.80&  1.46 \\
3.6828 & 20.60& 12.05&  0.21&  0.79&  1.45 \\
3.6832 & 20.66& 12.09&  0.22&  0.80&  1.46 \\
3.6836 & 21.52& 12.59&  0.22&  0.83&  1.52 \\
3.6844 & 20.61& 12.06&  0.22&  0.79&  1.45 \\
3.6850 & 21.44& 12.54&  0.22&  0.83&  1.51 \\
3.6855 & 19.98& 11.69&  0.21&  0.77&  1.41 \\
3.6863 & 19.79& 11.58&  0.21&  0.76&  1.40 \\
3.6867 & 19.92& 11.66&  0.21&  0.77&  1.41 \\
3.6875 & 19.88& 11.63&  0.21&  0.77&  1.41 \\
3.6882 & 20.42& 11.95&  0.21&  0.79&  1.44 \\
3.6886 & 20.19& 11.81&  0.21&  0.78&  1.43 \\
3.6893 & 25.36& 14.84&  0.26&  0.98&  1.79 \\
3.6908 & 20.97& 12.27&  0.22&  0.81&  1.49 \\
3.6939 & 23.11& 13.52&  0.24&  0.89&  1.64 \\
3.6979 & 22.38& 13.09&  0.23&  0.86&  1.59 \\
3.7017 & 21.89& 12.81&  0.23&  0.84&  1.56 \\
3.7068 & 22.56& 13.20&  0.24&  0.87&  1.62 \\ \hline \hline
\end{tabular}\\
\end{center}
\end{table}

\indent
For the selection of $\psip\ra\PPJP$ events,
we select a pair of low energy 
pions and determine the mass recoiling against 
these two pions, $m_{recoil}$, which shows a strong $\J$ peak, 
corresponding to the decay $\psip\ra\PPJP$. The $m_{recoil}$
distribution is fitted with a signal shape plus polynomial background 
to obtain the number of $\psip\ra\PPJP$ events at each energy point. The 
very clean $\psip\ra\PPJP$, $\J\ra l^+l^-$ channel is used to determine
the signal shape. The inclusive  $m_{recoil}$ distribution for all 24 
energy points combined is shown in Fig. \ref{m_recoil_incl}.  
For more detail, see Ref. \cite{fredppj}. 
The numbers of events selected for the four final states at the
24 energies are listed in Table~\ref{offres}.

\begin{table}[htb]
\doublerulesep 0.5pt
\caption{\label{offres} Numbers of events selected. }
\vskip 0.2 cm
\begin{center}
\begin{tabular}{c|rrrr}  \hline \hline
$E_{cm}$&\multicolumn{4}{|c}{ Selected number of events } \\ \cline{2-5}
  (GeV) &$N^{obs}_{ee}$
	         &$N^{obs}_{\mu \mu}$
	                  &$N^{obs}_h$
                                   &$N^{obs}_{\PPJP}$ \\ \hline \hline
  3.6668&  1789  &   110  &    385 &     7.2\\
  3.6719&  1752  &    92  &    389 &    10.2\\
  3.6750&  1734  &   124  &    391 &    22.3\\
  3.6781&  1783  &   102  &    437 &     0.0\\
  3.6801&  1771  &    66  &    485 &    12.8\\
  3.6809&  1909  &   104  &    588 &    14.2\\
  3.6820&  1790  &    91  &    764 &    75.0\\
  3.6828&  1773  &    86  &   1538 &   181.0\\
  3.6832&  1778  &    69  &   2662 &   426.1\\
  3.6836&  1852  &    79  &   4178 &   668.0\\
  3.6844&  1774  &    92  &   8169 &  1508.0\\
  3.6850&  1845  &   128  &  14684 &  2660.8\\
  3.6855&  1720  &   140  &  15073 &  2758.1\\
  3.6863&  1703  &   191  &  15621 &  2902.6\\
  3.6867&  1715  &   176  &  14980 &  2572.0\\
  3.6875&  1711  &   161  &  11658 &  2201.8\\
  3.6882&  1758  &   175  &   7047 &  1245.1\\
  3.6886&  1738  &   125  &   4960 &   868.9\\
  3.6893&  2183  &   188  &   3964 &   707.4\\
  3.6908&  1805  &   113  &   1897 &   269.0\\
  3.6939&  1989  &    97  &   1318 &   128.8\\
  3.6979&  1926  &   132  &   1029 &    82.6\\
  3.7017&  1884  &   118  &    876 &    77.2\\
  3.7068&  1942  &   120  &    729 &    26.5\\ \hline
   sum  &  43624 &   2879 &113823  & 19425.4\\ \hline \hline
\end{tabular}\\
\end{center}
\end{table}

\section{Acceptance}
The acceptance is the product of the trigger efficiency and
the reconstruction-selection
efficiency. The triggers are the same as those used in our $R$ scan
experiment \cite{trig},  The trigger efficiencies, measured by comparing the
responses to different trigger requirements in special runs taken at the
$\J$ resonance, are determined to be
1.000, 0.994 and 0.998 for $\EE,\MM$ and hadronic events respectively,
with an uncertainty of 0.005.

Different generators are used to determine the reconstruction-selection
efficiencies. For $\EE$ and $\MM$ final states, the efficiencies for QED
processes are determined in simulations with the BHABHA and MUPAIR
generators~\cite{gtor.1}. The resonance  $\EE$ and $\MM$ efficiencies are 
determined using the generator V2LL, adapted from MUPAIR, with 
the initial state radiative corrections removed. For the hadronic processes, 
an event generator for charmonium inclusive decay \cite{gtor.2a} is
used to obtain the efficiency for the resonance portion, and
the JETSET string fragmentation algorithm with parameters modified 
to fit the experimental data in the BEPC energy region~\cite{gtor.2b}
is used to compute the 
efficiency for the continuum portion. For the $\PPJP$
acceptance, a phase space Monte Carlo program, modified to give the correct dipion mass and angular
distributions \cite{pipijpsi}, is used.  

\begin{table}[thb]
\doublerulesep 0.7 pt
\caption{\label{selres} Acceptances for continuum, $A^c$, and the
resonance, $A^r$.}
\center
\begin{tabular}{ccccc}\hline \hline
final state    & $ \EE $ & $\MM $ & $had$     & $\PPJP$ \\ \hline
  $A^c (\%)$ & 72.4    & 37.1   & 74.5      & --     \\
$\delta A^c/A^c(\%)$
               &  3.2    &  5.0   &  7.1      & --     \\
  $A^r(\%)$ & 76.4    & 41.9   & 77.1      & 43.4   \\
$\delta A^r/A^r(\%)$
               &  8.5    &  4.7   &  2.2      &  3.4   \\ \hline \hline
\end{tabular} 
\end{table}

The acceptances of the four final states for continuum ($A^c$)
and resonance ($A^r$) processes, together with their relative errors,
are listed in Table~\ref{selres}. Here the acceptance for the $\EE$ final
state in the table applies to events within a restricted
solid angle ($ |\cos \theta_e | \leq 0.72 $),
whereas those of the hadron and dimuon final states cover all solid angles.
The acceptance error includes the uncertainties estimated by varying
selection cuts and using different selection methods and Monte Carlo models.

\section{Fit of observed cross sections and results}
The $\EE \rightarrow$ hadrons, $\pi^+ \pi^- J/\psi$,
$\EE$, and $\MM$ events at the 24 scan points
are fitted simultaneously to obtain 
the partial widths of $\psip$ to hadrons, $\pi^+ \pi^- J/\psi$, 
and $\MM$ final states. The total width is assumed to be the sum 
of four partial widths,
$\Gamma_t=\Gamma_{h}+\Gamma_{\mu}+\Gamma_{e}+\Gamma_{\tau}$, and
lepton universality is assumed\footnote{BES collaboration
has measured the branching fraction of $\psip$ decay into $\TT$.
This value along with those of the branching
fractions in $\EE$ and $\MM$, satisfies the relation predicted by
the sequential lepton hypothesis within errors\cite{brtt}.}, 
$ \Gamma_{e}=\Gamma_{\mu}=\Gamma_{\tau}/0.38847 $.
A likelihood function is constructed \cite{porter}.
Two kinds of correlations are taken into consideration in the
fitting formula: one is the correlation between different channels at
the same energy point, because
the same luminosity is used to determine the
cross sections at each energy; the other is the correlation
between different energy points for each channel, due to the acceptance
uncertainty being the same for all scan points.
The MINUIT package \cite{minuit} for maximization is used to give
the best estimates for $\psip$ parameters and their uncertainties.
The theoretical cross section used in the fit for the hadron channel
uses a Breit-Wigner amplitude and a non-resonant direct-channel amplitude 
plus a $\J$ resonance ``tail'' cross section, as determined by a previous 
BES $\J$ scan experiment \cite{jpsi}. The contributions from $\psip$ 
decay into $\EE$, $\MM$, $\TT$ final states mis-identified as hadron 
events have been taken into account by correcting the resonant 
hadron event acceptance\footnote{The contaminations from QED processes
$\EE \rightarrow l^+ l^-$ have been already subtracted, see 
section {\bf \ref{sect_two}}.}.
For the $\PPJP$ final state, 
only a resonant  Breit-Wigner amplitude is considered. In the $\MM$ channel, 
the $\psip$ resonant term, QED term, and their interference 
are included. The radiative corrections to these three processes are 
taken into account by the formulation of
Refs.~\cite{rad.1} and~\cite{rad.2}. For the $\EE$ final state,
where the QED $t$-channel photon exchange also contributes,
a theoretical cross section including radiative corrections is derived 
using the method of Ref.~\cite{rad.3}.
The effects on lepton-pair final state cross sections coming
from vacuum polarization are also taken into consideration~\cite{rad.4};
while for hadronic final states, the vacuum polarization is absorbed into
the definition of $\Gamma_{ee}$, that is

$$\Gamma_{exp}(\psip \rightarrow \EE) =$$
$$\Gamma_{0}(\psip \rightarrow \EE)
\cdot \frac{1}{|1-\prod (M^2_{\psip})|^2} , $$
where $\Gamma_{exp}$ is the experimental width, $\Gamma_{0}$ the lowest
order in $\alpha$ width, and $\prod$ the order $\alpha$ vacuum 
polarization~\cite{rad.5}.
The theoretical cross sections are convoluted with the energy distribution
of the colliding beams, which is treated as Gaussian. 
The following parameters are allowed to vary in the fit: 
the $\psi(2S)$ mass, $M$, the total width, $\Gamma_t$, the partial widths, 
$\Gamma_{\PPJP}$ and $\Gamma_l$, 
the energy spread of the machine, and the non-resonant hadronic cross section. 

As the branching fraction of $\psip$ to $\EE$ is small, the cross section
for the $\EE$ final state is dominated by the QED process. Therefore 
this channel is used to calculate the integrated luminosity at each 
energy point by an iterative method. 
First, all $\EE$ events within $|\cos\theta| \leq 0.72$ 
are taken as ``Bhabha'' events and used to calculate the
integrated luminosity at each energy point. A maximum likelihood fit is
performed to the observed cross sections for hadron, $\PPJP$, and muon pair
final states, and a group of $\psip$ parameters is obtained with the
assumption of $e$-$\mu$-$\tau$ universality. 
Then, separating the $\EE$ events into a QED part, a resonance part and 
their interference, the integrated luminosity for 
each energy point is recalculated using the QED part only, and the
fitting procedures are redone
to get new values for the $\psip$ parameters. The iterative process  
is repeated until the value of the integrated luminosity 
at each energy point is consistent for two successive iterations.  
  
\begin{figure}[htb]
\includegraphics[height=8.0 cm,width=7. cm]{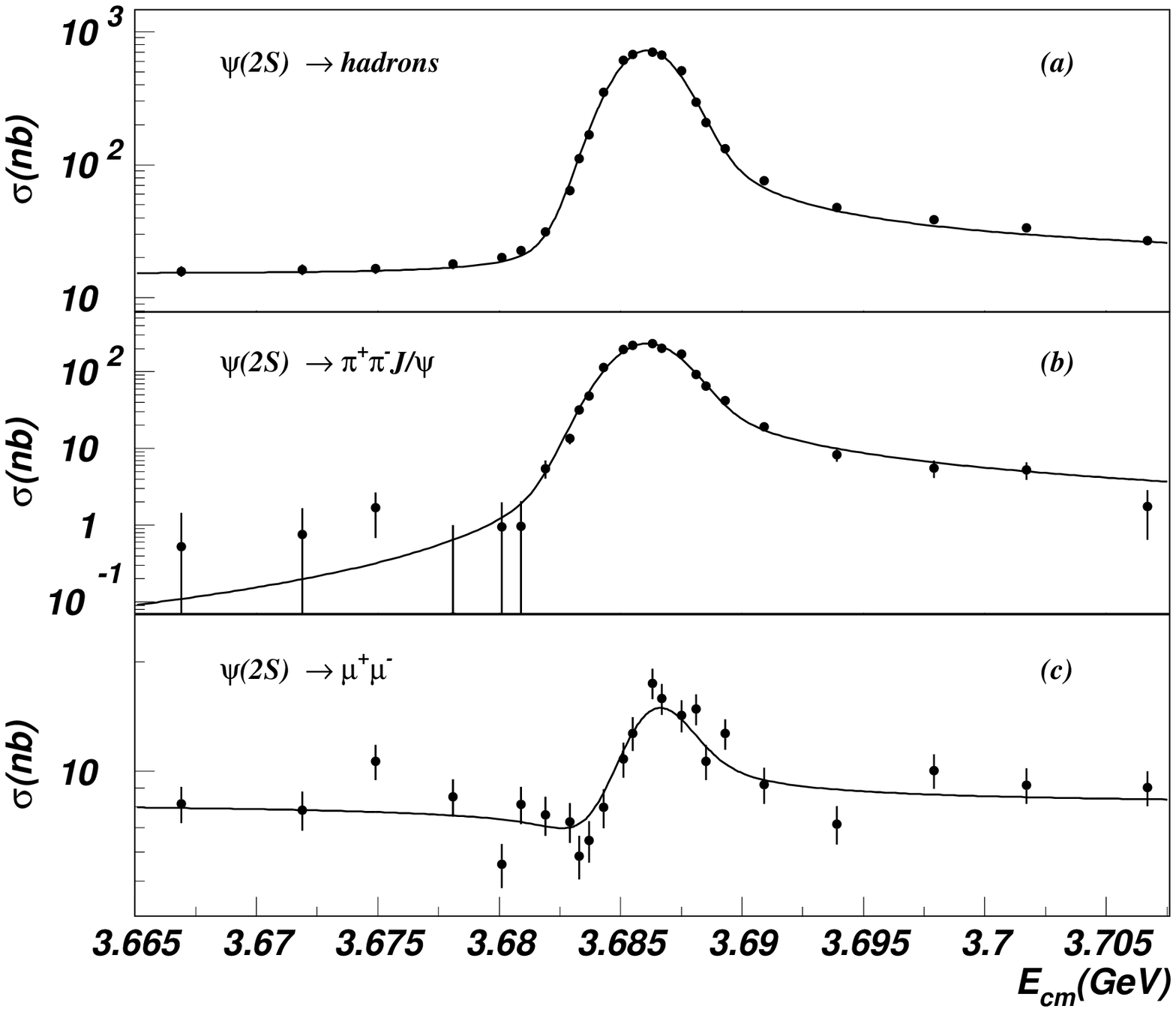}
\caption{The cross section for (a) $ e^+ e^- \rightarrow$ hadrons, (b)
$ e^+ e^- \rightarrow$ $\pi^+\pi^- J/\psi$, and (c)
$ e^+ e^- \rightarrow \mu^+ \mu^- $ versus center-of-mass energy. 
The solid curves represent the results of the fit to the data.}
\label{fig4c}
\end{figure}

The fitted curves are shown along with the scan points in Fig.~\ref{fig4c}. 
The fitted mass of the $\psip$ is corrected to the PDG value
\cite{PDG}.
The errors in the other parameters caused by this correction are negligible.
The fitted spread in the center-of-mass energy of the machine is
($1.298 \pm 0.007$) MeV,
in agreement with the expectation ($\sim$1.3 MeV). The resultant $R$
ratio for the  hadronic cross section near 
the $\psip$ resonance is $2.15 \pm 0.16$, which agrees well with 
the earlier BES $R$ measurements~\cite{R-scan}. 
 
\begin{table}[htb]
\doublerulesep 0.5pt
\caption{\label{fitres} Results and comparison with the PDG2002 \cite{PDG}.}
 \begin{tabular}{c||c|c} \hline \hline
      Value     &   BES        &    PDG2002             \\ \hline \hline
$\Gamma_t$(keV)$^{\dagger}$ 
                &$264\pm 27$   &$300 \pm 25$     \\ 
		&    (10.1\%)  &   (8.3\%)       \\ \hline
$\Gamma_h$(keV) &$258\pm 26$   &                 \\ 
		&    (10.1\%)  &                 \\ \hline 
$\Gamma_{\PPJP}$(keV)$^{\dagger}$
                &$85.4 \pm 8.7$&                 \\
		&    (10.1\%)  &                 \\ \hline
$\Gamma_l$(keV)$^{\dagger}$ 
                &$2.44\pm0.21$ & $2.19\pm0.15$   \\
		&    (8.8\%)   & (6.8\%)$^{\ddagger}$ \\ \hline
${\cal B}_{h}(\%)$
                &$97.79\pm0.15$&$98.10\pm0.30$   \\ 
         	&    (0.16\%)  &  (0.31 \%)      \\ \hline
${\cal B}_{\PPJP}(\%)$
                &$32.3\pm1.4 $ &$30.5\pm 1.6$    \\ 
		&    (4.4\%)   &  (5.2 \%)       \\ \hline
${\cal B}_l(\%)$
                &$0.93\pm0.08$ &$0.73\pm0.04$     \\ 
                &    (8.5\%)   &  (5.5\%)$^{\ddagger}$ \\ \hline \hline
				
\end{tabular} \\
{\footnotesize
Note : $\dagger$ indicates directly fitted values while others are
derived quantities; 
$\ddagger$ indicates the PDG values for $\Gamma_e$ and ${\cal B}_e$.
The numbers in parenthesis denote the relative errors. }
\end{table}

The results of the fit for decay widths and branching fractions are given
in Table~\ref{fitres}, together with corresponding PDG \cite{PDG} values for
comparison. The errors are the sum in quadrature of statistical,
fitting, and systematic uncertainties, including those from acceptance
uncertainties, luminosity uncertainties, 
and a center-of-mass energy uncertainty 
of 0.10 MeV~\cite{enun01}.
The systematic error for $\Gamma_t$, $\Gamma_{\PPJP}$ and $\Gamma_l$ 
is ($2.24 - 4.7$) \%, 3.2 \% and ($6.4 - 7.0$) \%, 
respectively. The correlation coefficients between directly fitted
parameters obtained from the fitting are utilized in the calculation of   
the errors of indirectly determined parameters such as $\Gamma_h$ and
branching fractions.
For ${\cal B}_{h}$, ${\cal B}_{\PPJP}$, and ${\cal B}_l$, 
the uncertainty related to the common error from the luminosity 
measurement cancels out. 

As a check of the fitting procedure, ${\cal B}_{\PPJP}$
was also determined by a simpler approach.  In this approach, the
distribution of $N^{obs}_h$ versus energy was fit with the shape
determined from $N^{obs}_{\PPJP}$ versus scan energy plus a
polynomial to represent the continuum process to determine the number
of hadrons coming from $\psi(2S)$ decays.  Using the ratio of the
total number of $N^{obs}_{\PPJP}$ events corrected by their
detection efficiency and the number of $\psi(2S)$ hadronic decays
corrected by their detection efficiency, we directly determine
${\cal B}_{\PPJP}$, which agrees very well with the result
from the full fitting procedure.

The assumption that lepton pairs couple to the $\psip$ only
via an intermediate photon~\cite{ref.3} implies the existence of the decay
$\psip\ra \gamma \rightarrow$ hadrons with a branching fraction:
$$
 \Gamma_{\gamma h}/\Gamma_t=R \Gamma_{\mu}/\Gamma_t=0.0199 \pm 0.0019,
$$
which corresponds to a width $\Gamma_{\gamma h}$ of $5.26\pm 0.32$ keV.
 
The $\psip$ lepton width $\Gamma_l=(2.44\pm0.21)$ keV obtained in this 
measurement agrees with the BES previous value of 
$\Gamma_e=(2.07\pm0.32)$ keV~\cite{gaee}, within the error. 
In addition, 
this is the first direct measurement to the decay width 
of $\psip\ra\PPJP$, and the precision of ${\cal B}_{\PPJP}$ is much better than
previous measurements and the current PDG value.\\

\section*{Acknowledgments }

We gratefully acknowledge the efforts of the staffs of the BEPC 
accelerator and the computing center at the Institute of High Energy 
Physics, Beijing. This work is supported in part by 
the National Natural Science Foundation
of China under contracts Nos. 19991480,10175060
and the Chinese Academy of Sciences under contract No. KJ 95T-03(IHEP); and
by the Department of Energy under Contract Nos.
DE-FG03-92ER40701 (Caltech), DE-FG03-93ER40788 (Colorado State University),
DE-AC03-76SF00515 (SLAC), DE-FG03-94ER40833 (U Hawaii), 
DE-FG03-95ER40925 (UT Dallas).



\begin{thebibliography}{99}
\bibitem{track_cuts}
The track momenta are required to 
satisfy the following requirements:

\[\begin{array}{l}
\tilde{p}_1 \geq 0.95, \mbox{\hskip 0.2cm  or \hskip 0.2cm}
\tilde{p}_2 \geq 0.95, \mbox{\hskip 0.2cm  or \hskip 0.2cm}\\
\sqrt{(\tilde{p}_1-1)^2+(\tilde{p}_2-1)^2} < 0.16,
\end{array} \]
for an $e$-pair, where $\tilde{p}$ is the normalized momentum:
$\tilde{p}=p/E_{beam}$. For a $\mu$-pair, the corresponding cut is:
$$
 \makebox[1.5cm]{$P$-$cut.1$} \cap
(\makebox[1.5cm]{$P$-$cut.2$} \cup
\makebox[1.5cm]{$P$-$cut.3$}) $$
where $P$-$cut.1$, $P$-$cut.2$, and $P$-$cut.3$ are defined as following:
\begin{eqnarray*}
 \makebox[1.cm]{$P$-$cut.1$:}&
 \sqrt{\displaystyle
 \left(\frac{\tilde{p}_1-\tilde{p}_2}{0.35}\right)^2+
      \left(\frac{\tilde{p}_1+\tilde{p}_2-1.68}{0.125}\right)^2} >1, \\
 \makebox[1.cm]{$P$-$cut.2$:}& \mbox{} \hskip -2.5cm
 (\tilde{p}_1 \geq 0.9) \cap (\tilde{p}_2 \geq 0.7),  \\
 \makebox[1.cm]{$P$-$cut.3$:}& \mbox{} \hskip -2.5cm
 (\tilde{p}_2 \geq 0.9) \cap (\tilde{p}_2 \geq 0.7).
\end{eqnarray*}
These momentum cuts are shown in Fig.~\ref{mixpcut}.
\end{thebibliography}
\end{document}